\documentclass[reprint,  amsmath, amssymb,  aps, prd, nofootinbib, superscriptaddress]{revtex4-2}
\usepackage{indentfirst}
\usepackage{graphicx}
\usepackage{dcolumn}
\usepackage{bm}
\usepackage{amsmath}
\usepackage{amsfonts}
\usepackage{amssymb}
\usepackage{physics}
\usepackage{braket}
\usepackage{color}
\usepackage{autobreak}
\usepackage{hyperref}
\hypersetup{colorlinks=true, citecolor=blue, urlcolor=blue, linkcolor=blue}

\begin{document}
\title{Current-induced inverse symmetry breaking and asymmetric critical phenomena at current-driven tricritical point}
\author{Masataka Matsumoto}
\affiliation{Department of Mathematics, Shanghai University, Shanghai 200444, China}
\author{Shin Nakamura}
\affiliation{Department of Physics, Chuo University, Tokyo 112-8551, Japan}

\begin{abstract} 
We study critical phenomena associated with a spontaneous chiral symmetry breaking in current-driven non-equilibrium steady states by using holography.
We find that the critical exponents $(\gamma, \nu)$ at the tricritical point are asymmetric between the chiral symmetry restored phase and the broken phase. Their values in the broken phase are different from those of the mean-field theory, whereas other critical exponents are the mean-field values.
The phase diagram with respect to temperature and current density shows a re-entrant structure: the broken chiral symmetry is restored again at low temperatures in the presence of current density.
\end{abstract}

\maketitle

\section{Introduction}
Non-equilibrium steady state (NESS) with a constant flow of current is a natural extension of equilibrium states to out of equilibrium. 
Investigation of the role of the current in phase transitions in NESS is an important challenge\,\footnote{For recent studies on phase transitions with heat current, see, for example, \cite{Sasa-Nakagawa} and the references therein}. 

We study current-driven critical phenomena of NESSs whose microscopic theory is explicitly defined by quantum gauge theory.
We employ the gauge/gravity duality (holography)
\,\cite{Maldacena:1998,Gubser:1998,Witten:1998}. Various applications of the gauge/gravity duality to non-equilibrium physics have been reported (for example, see reviews\,\cite{Hubeny:2010ry,Kundu:2019ull}). Recently, the authors have found a phase transition associated with a spontaneous chiral symmetry breaking in a current-driven system\,\cite{Imaizumi:2019byu}. The phase diagram contains both the tricritical point (TCP) and the critical line (CL) which appears in the presence of the current. However, the whole phase structure and the critical phenomena have not been investigated.

In this paper, we report an ``inverse symmetry breaking" and asymmetric critical phenomena at the TCP in the current-driven system. 
The inverse symmetry breaking is the spontaneous symmetry breaking at higher temperatures, rather than at lower temperatures.
An example of this counterintuitive phenomenon is the inverse melting of materials.
The inverse symmetry breaking is also discussed in the context of early universe. 
See reviews\,\cite{Bajc:1999cn,Senjanovic:1998xc}.

The asymmetric critical phenomena are the critical phenomena whose critical exponents are different depending on whether we approach the critical value from the lower values of the control parameter or the higher values.
We find that the critical phenomena in our system are asymmetric only at the TCP. An example of asymmetric critical behavior has been found in the Sachdev-Ye-Kitaev model\,\cite{Ferrari:2019ogc,Cao:2021upq}. However, to the best knowledge of the authors, asymmetric critical phenomena that appear only at TCP have not been reported elsewhere. Interestingly, the asymmetric critical exponents are not the mean-field values\,\footnote{In this paper, we use the term ``mean-field values" to
mean the values of critical exponents derived from the conventional classical Landau theory.} in spite that we are taking the large-$N$ limit.

\section{Setup}
We consider a strongly-coupled $SU(N)$ ${{\cal N} =4}$ supersymmetric Yang-Mills (SYM) theory with ${{\cal N} =2}$ hypermultiplet in the large-$N$ limit. We set the mass of the hypermultiplet to zero so that we have a chiral symmetry at the level of the Lagrangian.
The particles in the ${{\cal N} =2}$ hypermultiplet carry a global $U(1)$ charge which we call ``electric charge'' in this paper. When we apply an external electric field, they form a NESS with finite current density. The ${{\cal N} =4}$ SYM sector plays the role of a heat bath.
We set the charge density of the system to zero: we have an equal number of positively charged particles and negatively charged particles. We also apply an external magnetic field perpendicular to the external electric field.
In this paper, we study phase transitions where the chiral symmetry is spontaneously broken at finite current density in the presence of the electromagnetic field.

The holographic dual of our system is the D3-D7 model\,\cite{Karch:2002sh} with an electric and magnetic field\,\cite{Karch:2007pd,Ammon:2009jt} in the probe limit. 
The dual geometry is the five-dimensional AdS-Schwarzschild black-hole times $S^{5}$:
\begin{equation}
    ds^{2}= \frac{L^{2}}{u^{2}}\left( -f(u)dt^{2} +\frac{du^{2}}{f(u)} +d\vec{x}^{2} \right)+L^{2}d\Omega^{2}_{5},
\end{equation}
where $f(u)=1-u^{4}/u_{\rm H}^{4}$. $u\,\left( 0 \leq u \leq u_{\rm H}\right)$ is the radial direction, $t$ and $\vec{x}=(x, y, z)$ are the coordinates for the (3+1)-dimensional spacetime of the gauge theory.
The black hole horizon is located at $u=u_{\rm H}$ and the boundary is located at $u=0$. The Hawking temperature is given by $T=1/(\pi u_{\rm H})$, which corresponds to the temperature of the heat bath.
$    d\Omega_{5}^{2}=d\theta^{2}+\sin^{2} \theta d\psi^{2} +\cos^{2}\theta d\Omega_{3}^{2},
$
where $d\Omega_{3}$ is the line element of the unit $S^{3}$. The D7-brane wraps the $S^{3}$ part.

The dynamics of the D7-brane is governed by the Dirac-Born-Infeld\,(DBI) action
\begin{equation}
	S_{\rm DBI} = -T_{\rm D7}\int d^{8}\xi \sqrt{-\det\left(g_{ab} +\left(2\pi l_{\rm s}^{2}\right) F_{ab} \right)},
\end{equation}
where $g_{ab}$ is the induced metric and $F_{ab}= \partial_{a}A_{b}-\partial_{b}A_{a}$ is the field strength of the ${U(1)}$ gauge field $A_{a}$ on the D7-brane. 
$T_{D7}$ is the tension of the D7-brane given by $T_{\rm D7}^{-1}=(2\pi)^{7}l_{\rm s}^{8}g_{\rm s}$, where $l_{\rm s}$ and $g_{\rm s}$ are the string length and the string coupling constant, respectively.
We set $L=1$ and $\left(2\pi l_{\rm s}^{2}\right)=1$, for simplicity. This corresponds to setting $2\lambda=(2\pi)^{2}$ where $\lambda=g_{\rm YM}^2 N$ is the 't Hooft coupling of the gauge theory.

The configuration of the D7-brane is detemined by $\theta(u)\,\left( 0\leq\theta\leq\pi/2 \right)$ and $\psi$. In our study, we take $\psi=0$ without loss of generality.
In addition, we apply an electric field in the $x$ direction and a magnetic field in the $z$ direction.
We employ the following ansatz for the gauge fields:\,$A_{x}(t,u)=-Et+h(u)$, $A_{y}(x)=Bx$, where $E$ and $B$ correspond to the electric field and magnetic field acting on the charged particles, respectively. $\theta(u)$ and $h(u)$ are expanded as the following asymptotic form near the boundary
$
\theta(u)= m u + \theta_{2} u^{3} + \cdots, \hspace{1em} h(u)= 
	Ju^{2}/2{\cal N}+ \cdots,
	\label{eq:asymp}
$
where $m$ and $J$ correspond to the mass of the charged particles and the electric current density.
Here, we define ${\cal N}= T_{\rm D7}(2\pi^{2})$. Note that the current is in the $x$ direction even in the presence of the magnetic field since the total charge density is zero. The operator conjugate to $m$ is the chiral condensate given by $\expval{\bar{q}q} = {\cal N} (2\theta_{2}-m^{3}/6)$\,\cite{Karch:2007pd}. In the presence of the current density $J$, the {\it effective horizon} emerges on the D7-brane outside of the black hole horizon. The effective horizon is a causal boundary for the modes governed by the open-string metric on the worldvolume of the D7-brane\,\cite{Kim:2011qh,Sonner:2012if,Nakamura:2013yqa}. The location of the effective horizon $u_{*}$ is determined by
$
    \left. B^{2}g_{tt} + E^{2}g_{xx}+g_{tt}g_{xx}^{2}\right|_{u=u_{*}}=0.
    \label{eq:ustar}
$
Then, 
$	J = \left. {\cal N} \sqrt{-g_{tt}}g_{xx}\cos^{3}\theta \right|_{u=u_{*}},
	\label{eq:current}
$
assuming the action remains real at any $u$. 
The details are discussed in \cite{Ammon:2009jt}.
Hereafter, we set ${\cal N}=1$ for simplicity.

In this paper, we consider the current density $J$ as a control parameter in the NESS system. Using the scale invariance of the system, we have two dimension-less parameters ($T/B^{1/2}$, $J/B^{3/2}$).

\section{Phase diagram}
We solve the equation of motion for $\theta(u)$ with the following boundary conditions. At the boundary, we impose the massless condition $m= \left. u^{-1} \theta(u)\right|_{u=0}=0$. 
At the effective horizon, we impose the regularity of $\theta(u)$ there\,\footnote{When the D7-brane does not reach the effective horizon, we impose the regularity at the point of the maximum of $u$. However, the D7-brane reaches the effective horizon at finite $J$, which we consider in this paper.}.

We find two types of solutions, namely $\theta=0$ and $\theta\neq0$. The former solution, which shows the chiral condensate vanishes\,($\expval{\bar{q}q}=0$), holds regardless of $E$ and $B$. The latter solution with non-vanishing chiral condensate is possible only in the presence of $B$\,\footnote{A similar solution that reaches the black hole horizon was obtained in the presence of $B$ by switching on the charge density instead of the electric current\,\cite{Evans:2011tk}.}. 
These two types of solutions correspond to the chiral symmetry restored ($\chi{\rm SR}$) phase and the chiral symmetry broken phase ($\chi{\rm SB}$), respectively\,\cite{Babington:2003vm}.  
The order parameter is the chiral condensate $\expval{\bar{q}q}$. 

\begin{figure}[tbp]
\centering
\includegraphics[width=7.5cm]{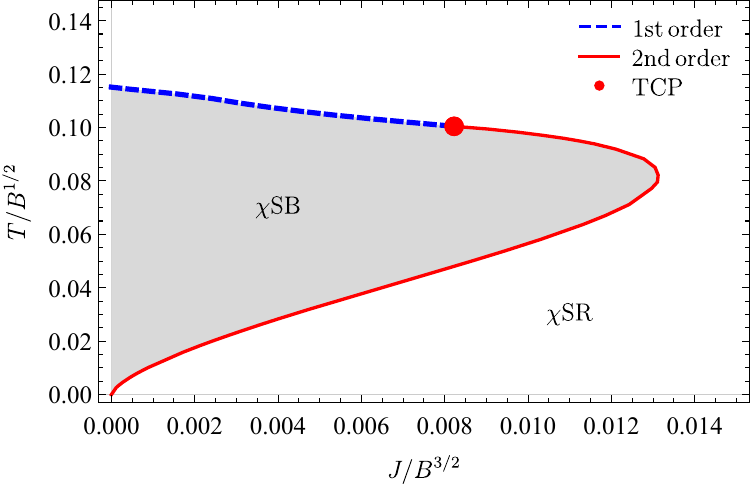}
\caption{Phase diagram with respect to $T/B^{1/2}$ and $J/B^{3/2}$. The gray shaded region corresponds the chiral symmetry broken\,($\chi {\rm SB}$) phase and the other region corresponds to the chiral symmetry restored\,($\chi{\rm SR}$) phase. The blue dashed curve denotes the first-order phase transition line and the red solid curve denotes the critical line. 
The red dot corresponds to the tricritical point\,\cite{Imaizumi:2019byu}.}
\label{fig:phase}
\end{figure}

We show the phase diagram with respect to ($T/B^{1/2}$, $J/B^{3/2}$) in Fig.\,\ref{fig:phase}.
The phase boundary is separated into the first-order phase transition line and the critical line (CL) where the second-order phase transition occurs. The point between them is the TCP\,\cite{Imaizumi:2019byu}.  As can be seen from Fig.\,\ref{fig:phase}, the CL and the TCP appear only at finite $J$. 
In the $\chi {\rm SB}$ phase, we suppose the solutions of $\theta\neq0$ are more stable than those of $\theta=0$ since the Hamiltonian density in the gravity side is smaller\,\cite{Imaizumi:2019byu}.
In the phase diagram, we find a re-entrant structure at small $T/B^{1/2}$. 
Compared to the phase diagram with respect to the charge density instead of the current density\,\cite{Evans:2011tk}, the re-entrant behavior is characteristic of the current-driven system.
We also numerically find that the CL at small $T/B^{1/2}$ and $J/B^{3/2}$ is well described by
\begin{equation}
	\left( \frac{T_{c}}{B_{c}^{1/2}}\right)^{\kappa} \approx 1.2\left(\frac{J_{c}}{B_{c}^{3/2}} \right),
\label{eq:12}
\end{equation}
where $\kappa\approx 3/2$ which is independent of $\lambda$ and $N$, whereas the factor $1.2$ depends on them.

\section{Critical phenomena}
In this paper, we compute the critical exponents $(\gamma, \nu, \eta, z)$ and $\delta$. We do not attempt to define the critical exponent $\alpha$, since the notion of heat capacity is not well-defined in NESS. 
The obtained values of $\delta$ are $3$ for the CL and $5$ for the TCP. We will give the details of the computation of $\delta$ in Appendix.  The values of $\beta$ have been obtained to be $1/2$ for the CL and $1/4$ for the TCP in\,\cite{Imaizumi:2019byu}. 

For computation of $(\gamma, \nu, \eta, z)$, we consider a small perturbation of $\theta$ that corresponds to a fluctuation of the order parameter. We assume that the perturbation can be written as $\delta\theta(t,u,\vec{x})=\vartheta(u) e^{-i \omega t + i k z}$ with the momentum $\vec{k}=(0,0,k)$ in $z$ direction, for simplicity.  If we choose the momentum in another direction, the equation of motion for $\delta\theta$ becomes more complicated.
In the $\chi{\rm SR}$ phase, where $\theta=0$, $\delta\theta$ does not couple to the other modes in the linear order because the background solution is trivial.
In the $\chi {\rm SB}$ phase with $\theta\neq0$, on the other hand, the perturbation couples to the fluctuation of the $x$-component of the gauge field $\delta A_{x}(t,u,\vec{x})$ that carries momentum in $z$ direction.
As a result, we obtain the equations of motion for the perturbations in each phase as explicitly shown in Appendix.
Near the boundary, the perturbation field for $\theta$ can be written as the following asymptotic form $\vartheta(u) = \vartheta^{(0)} u + \vartheta^{(1)} u^{3} + \cdots$,
where $\vartheta^{(0)}$ and $\vartheta^{(1)}$ are the non-normalizable mode and the normalizable mode, respectively. Following the analysis of critical phenomena in \cite{Maeda:2009wv}, we assume that each mode can be expanded as a function of $(\omega,k)$,
$\vartheta^{(0)}\sim \vartheta^{(0)}_{0} + \omega \vartheta^{(0)}_{(1,0)}+ k^{2} \vartheta^{(0)}_{(0,1)}$, $	\vartheta^{(1)}\sim \vartheta^{(1)}_{0} + \omega \vartheta^{(1)}_{(1,0)}+ k^{2} \vartheta^{(1)}_{(0,1)}$.
Using these expressions, the retarded Green's function $G^{R}_{\vartheta \vartheta}(\omega, k)$ is proportional to
\begin{eqnarray}
	\frac{\vartheta^{(1)}_{0} + \omega \vartheta^{(1)}_{(1,0)}+ k^{2} \vartheta^{(1)}_{(0,1)}}{\vartheta^{(0)}_{0} + \omega \vartheta^{(0)}_{(1,0)}+ k^{2} \vartheta^{(0)}_{(0,1)}}  
	 \sim 
\frac{\vartheta^{(1)}_{0}/\vartheta^{(0)}_{(0,1)}}{-ic\omega + k^{2} +1/\xi^{2}},
	 \label{eq:Green}
\end{eqnarray}
where $c\equiv i \vartheta^{(0)}_{(1,0)}/\vartheta^{(0)}_{(0,1)}$, and $\xi\equiv \sqrt{\vartheta^{(0)}_{(0,1)}/\vartheta^{(0)}_{0}}$ is the correlation length. If we take $\omega\to 0$, $G^{R}_{\vartheta \vartheta}(k) \propto 1/(k^{2}+1/\xi^{2})$. Then, if we find the poles of the retarded Green's function near the CL and the TCP, we can explore the behavior of the correlation length. 

We employ two definitions for $(\gamma, \nu)$: $(\gamma_{+}, \nu_{+})$ are those when we approach the CL and the TCP from the $\chi{\rm SR}$ phase, whereas $(\gamma_{-}, \nu_{-})$ are those when we approach them from the $\chi{\rm SB}$ phase. 
We define the critical exponent $\nu_{\pm}$ by using $J$ as
\begin{equation}
	\xi\propto \left|J-J_{c} \right|^{-\nu_{\pm}},
\end{equation}
where $J_{c}$ is the critical value of the current density. Note that we study critical phenomena with $T/B^{1/2}$ fixed. 
If we take the limit of $k \to 0$, the retarded Green's function agrees with the homogeneous susceptibility:\,$\chi \equiv \vartheta^{(1)}_{0}/\vartheta^{(0)}_{0}$, which is called the {\it chiral susceptibility} in terms of QCD. Thus, we define another critical exponent $\gamma_{\pm}$ as
\begin{equation}
	 \chi \propto \left| J-J_{c}\right|^{-\gamma_{\pm}}.
\end{equation}
If we take $\omega\to 0$ in (\ref{eq:Green}) at CL or TCP, we have $G^{R}_{\vartheta \vartheta}(k) \propto k^{\eta-2}$, where $\eta$ is  
{\it anomalous dimension}. 
In (\ref{eq:Green}), we have assumed $\eta=0$ as numerically confirmed later. Note that one can see that the scaling relation for the Green's function $\gamma_{\pm}=\nu_{\pm}(2-\eta)$ is satisfied for the above three critical exponents.
We can determine the dynamic critical exponent $z$ defined by
$\tau_{k=0}\propto \xi^{z}$,
where $\tau_{k=0}$ is the relaxation time of a homogeneous perturbation. 
We can determine the dynamic critical exponent from the critical dispersion relation $\omega \propto k^{z}$ 
at the CL and the TCP. In (\ref{eq:Green}), we have assumed that $z=2$, which we will confirm numerically later.

\begin{figure}[t!bp]
\centering
\includegraphics[width=7.5cm]{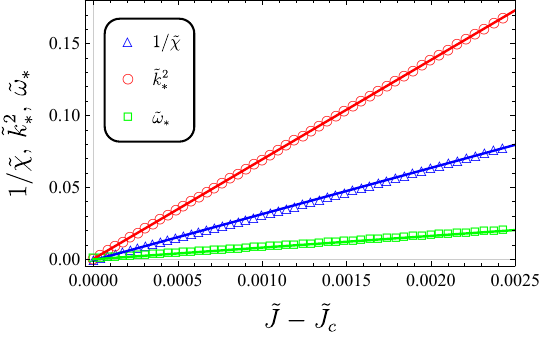}
\includegraphics[width=7.5cm]{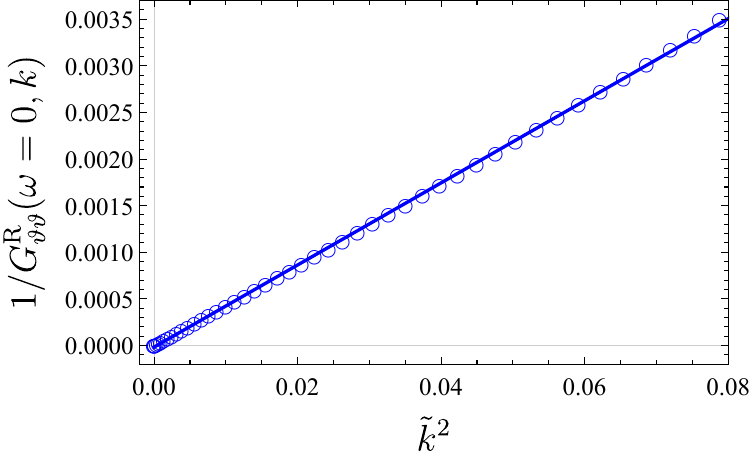}
\caption{Critical behaviors at a CP in the $\chi {\rm SR}$ phase at $T/B^{1/2}\approx 0.096$. 
}
\label{fig:CP}
\end{figure}

Fig.\,\ref{fig:CP} shows the typical critical behaviors at a critical point (CP) in the $\chi {\rm SR}$ phase.
The tilde denotes the scaled dimensionless quantities, such as $\tilde{\chi}=\chi/B^{1/2}$, $\tilde{J}=J/B^{3/2}$, $\tilde{k}=k/B^{1/2}$ and $\tilde{\omega}=\omega/B^{1/2}$. $\tilde{k}_{*}$ and $\tilde{\omega}_{*}$ represent the location of the pole $(\tilde{\omega},\tilde{k})=(0,\tilde{k}_{*})$ or $(\tilde{\omega},\tilde{k})=(\tilde{\omega}_{*},0)$ of the retarded Green's function given by (\ref{eq:Green}) at given value of $\tilde{J}-\tilde{J}_{c}$.
Note that the imaginary part of $\tilde{\omega}_{*}$ is inversely proportional to $\tau_{k=0}$.
These numerical results show that $(\gamma_{+}, \nu_{+}, \eta, z)=(1, 1/2, 0, 2)$.
They are the same as the critical exponents of the mean-field theory for equilibrium phase transitions.
Note that $z=2-\eta$ is corresponding to a model with non-conserved order parameter\,\cite{RevModPhys.49.435}. This is consistent with the fact that the chiral condensate is not a conserved quantity. We confirm that the critical behaviors at the CL in the $\chi{\rm SB}$ phase also show the same values of the critical exponents as explicitly shown in Appendix.

\begin{figure}[tbp]
\centering
\includegraphics[width=7.5cm]{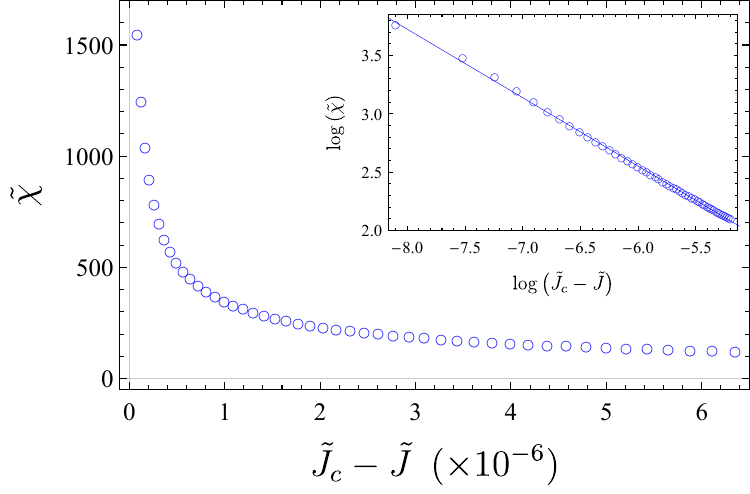}
\includegraphics[width=7.5cm]{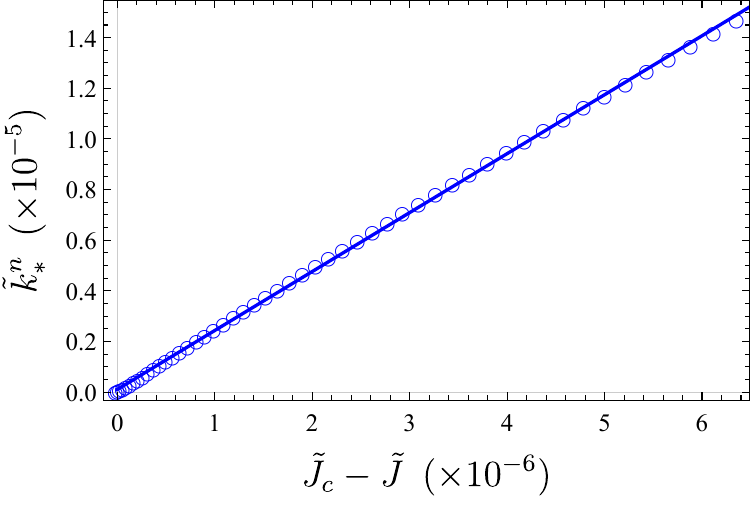}
\caption{The critical behaviors of $\tilde{\chi}$\,(top) and $\tilde{k}_{*}$\,(bottom) at TCP in the $\chi{\rm SB}$ phase.
The inset of the top panel shows the log-log plot from which we obtain $\gamma_{-}\approx 0.58$. In the bottom panel, $n\approx 0.29^{-1}$.}
\label{fig:TCP}
\end{figure}

At the TCP, we again obtain the mean-field values 
$(\gamma_{+}, \nu_{+}, \eta, z)=(1, 1/2, 0, 2)$
in the $\chi{\rm SR}$ phase. On the other hand, we obtain different critical behaviors for $(\gamma_{-},\nu_{-})$ in the $\chi{\rm SB}$ phase. In Fig.\,\ref{fig:TCP}, we show the critical behaviors of $\tilde{\chi}$ and $\tilde{k}_{*}^{n}$ with $n\approx 0.29^{-1}$ as a function of $\tilde{J}_{c}-\tilde{J}$.
Our numerical results imply that the critical exponents $(\gamma_{-},\nu_{-})$ are approximately given by $(0.58,0.29)$ at the TCP in the $\chi{\rm SB}$ phase. These values of the critical exponents are obviously different from the mean-field values. Note that these values also satisfy the scaling relation for the Green's function $\gamma_{-}=\nu_{-}(2-\eta)$ since we confirm that $\eta=0$ at the TCP.
On the other hand, these values do not satisfy 
$\gamma = \beta(\delta-1)$ that comes from the scaling hypothesis for the free energy. Violation of this scaling relation at TCP in the $\chi{\rm SB}$ phase is also a novel feature of our results in this paper.

\begin{figure}[tbp]
\centering
\includegraphics[width=7.5cm]{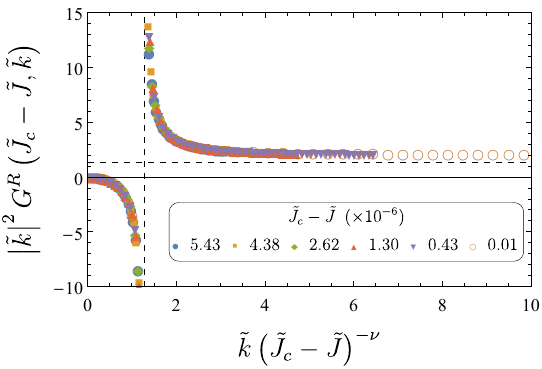}
\includegraphics[width=7.5cm]{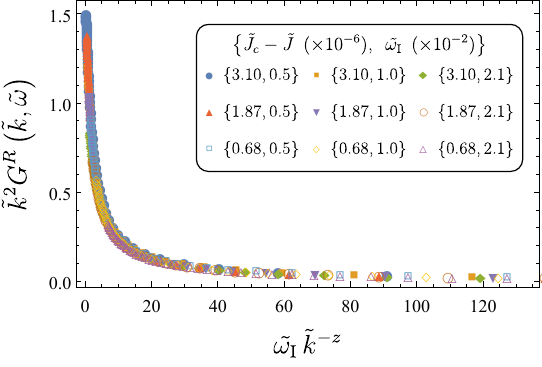}
\caption{The top panel shows the scaling function $\hat{g}_{\pm}$ as a function of $\tilde{k}(\tilde{J}_{c}-\tilde{J} )^{-\nu}$ with $\omega=0$. The dashed lines denote the position of the poles. The bottom panel shows the scaling function $\hat{g}_{\pm}$ as a function of $\tilde{\omega}_{\rm I}\tilde{k}^{-z}$ with $\tilde{k}(\tilde{J}_{c}-\tilde{J} )^{-\nu}$ fixed, where $\tilde{\omega}_{\rm I}$ is the imaginary part of $\tilde{\omega}$. We choose the best fitting parameter $\nu\approx 0.29$ and $z \approx 2$.}
\label{fig:scale}
\end{figure}

To corroborate the peculiar values of $(\gamma_{-},\nu_{-})$, we also perform the dynamic scaling analysis. Here, we apply the dynamic scaling hypothesis to our system (for example, see\,\cite{Tuber2014CriticalDA}). Then, the retarded Green's function can be written as the following scaling form
\begin{equation}
	G^{R}_{\vartheta \vartheta}(t,k,\omega)= \abs{k}^{-2+\eta} \hat g_{\pm}( k t^{-\nu_{\pm}}, \omega k^{-z}),
	\label{eq:scale}
\end{equation}
where $t$ represents the deviation from a critical point. In our case, we define $t=|\tilde{J}-\tilde{J}_{c}|$. 
Here, $\hat{g}_{\pm}$ are some functions whose $\pm$ represents whether we are in the $\chi{\rm SR}$ phase $(+)$ or in the $\chi{\rm SB}$ phase $(-)$. 
Using the scaling form, $\abs{k}^{2-\eta}G^{R}_{\vartheta \vartheta}$ can be described as a function of $kt^{-\nu_{-}}$ when $\tilde{\omega}=0$. 
In Fig.\,\ref{fig:scale}, we plot $\abs{k}^{2}G^{R}_{\vartheta \vartheta}$ in such a way that all the plots are on a single curve at various combinations of the parameters.
From the top panel of Fig.\,\ref{fig:scale}, we obtain $\nu\approx 0.29$. In the bottom panel of Fig.\,\ref{fig:scale},
we find that all the plots are on a single curve if we choose $z=2$ and $\nu=0.29$. Note that we have substituted $\eta=0$.

\section{Discussion}
In this paper, we have studied the critical phenomena of NESS system driven by the constant current flow. One of the essential differences from the phase diagram in equilibrium\,\cite{Evans:2011tk} is the re-entrant structure. 
The re-entrant phase structure indicates that the current-induced symmetry restoration is enhanced at small temperatures\,\footnote{It might be worthwhile to compare the field-theory analysis of inverse symmetry breaking of equilibrium systems given in\,\cite{Farias:2021ult}. The authors of \cite{Farias:2021ult} state that the temperature of the inverse symmetry breaking grows in the presence of a magnetic field. If we fix the current density, the magnetic field raises the critical temperature in our system, too.}.
It has been known that the re-entrant behavior emerges owing to the presence of disorders in simple models\,\cite{Crisanti2005,Thomas2011}. 
It is interesting to study whether the current plays a role of disorder in our system.

We have shown that the critical exponents ($\gamma_{\pm},\nu_{\pm},\eta,z$) at the CL agree with those in the mean-field theory.
At the TCP, on the other hand, we have found that ($\gamma_{\pm},\nu_{\pm}$) are asymmetric between the $\chi{\rm SR}$ phase and the $\chi{\rm SB}$ phase. 
Our results imply that the critical phenomena of our phase transitions at the CL can be formulated by the Landau theory\,\footnote{Our previous studies also imply that there could be a Landau-like theory even in the NESS regime\,\cite{Nakamura:2012ae,Imaizumi:2019byu,Matsumoto:2018ukk}.}, whereas those at the TCP are quite different from the conventional Landau theory. To best our knowledge, an asymmetric critical behavior only at a TCP is observed for the first time. It would be interesting to investigate whether the asymmetry we have discovered are characteristic of non-equilibrium systems or not.

One of the possibilities is the contribution of a dangerously irrelevant variable near the TCP\,\cite{PhysRevB.13.2222,PhysRevLett.115.200601}. The violation of $\gamma = \beta(\delta-1)$ at TCP in the $\chi$SB phase suggests that the conventional scaling hypothesis does not work there. We leave the investigation of these issues for future work.

We expect that the novel phenomena found in the present work
can be experimentally detectable in a system with gapless chiral fermions such as the Dirac semimetals in the presence of the electric field and the magnetic field.

\begin{acknowledgements}
The authors are grateful to N.\,Sogabe for helpful discussions. The authors also thank RIKEN iTHEMS NEW working group for fruitful discussions. The work of M.\,M.\, is supported by National Natural Science Foundation of China Grant No.\,12047538. The work of S.\,N.\, is supported in part by JSPS KAKENHI Grants No.\,JP19K03659, No. JP19H05821, and the Chuo University Personal Research Grant. 

\end{acknowledgements}

\appendix
\section{Perturbations}
The equation of motion for $\delta \theta=\vartheta(u)e^{-i\omega t + i k z}$ on the trivial background $\theta=0$ is given by
\begin{widetext}
\begin{align}
\begin{autobreak}
 F \vartheta''
 +\left[2 i E \omega  u^4 h' +\frac{u^3 f h'^2 \left(u f'-6 f\right)}{2} -\frac{F+ 2f}{u}-\frac{f'}{2 f}F \right]\vartheta'
 +\Biggl[ \Biggr. \Bigl(3+\left(\omega^{2}-k^{2}\right)u^{2} \Bigr)u^{2}h'^{2}+ \frac{\left(3-k^{2}u^{2}\right)}{u^2 f} F
 +\frac{\omega ^2 \left(B^2 u^4+1\right)}{f}
 -\frac{i E \omega  u^3 h' \left(u f'-2 f\right)}{f} \Biggl.\Biggr]\vartheta=0,
\end{autobreak}
 \label{eq:eom0}
\end{align}
\end{widetext}
where $F(u)=\left(B^{2}u^{4}+1 \right)f(u)-E^{2}u^{4}$.
Here, the prime denotes the derivative with respect to $u$. $h^{\prime}$ that appears in (\ref{eq:eom0}) is given by 
\begin{equation}
	h'(u)^{2} = -\frac{J^{2}u^{2}F}{\left(J^{2}u^{6}-f\right)f^{2}}.
 \label{eq:hprime}
\end{equation}
The location of the effective horizon is determined by $F(u_{*})=0$.
Note that the current density $J$ is given so that both the numerator and the denominator of (\ref{eq:hprime})  simultaneously become zero at $u=u_{*}$.

On the nontrivial background of $\theta\neq 0$, $\delta\theta$ couples to the fluctuation of the $x$-component of the gauge field $\delta A_{x}=a(u)e^{-i\omega t + i k z}$.
The equations of motion for these perturbations are given by
\begin{eqnarray}
	\vartheta'' + A\vartheta' +Ba' + C\vartheta + D a&=&0, \\
	a'' + \tilde{A}\vartheta' +\tilde{B}a' + \tilde{C}\vartheta + \tilde{D} a&=&0,
\end{eqnarray}
where
\begin{widetext}
\begin{align}
\begin{autobreak}
	A=
	\frac{1}{2uF} \Biggl[ \Biggr. 12ufF\theta' \tan \theta +3u^{2}f^{2}\theta'^{2}\left( uF'-8F\right) +u^{4}f^{2}h'^{2}\left( uf'-6f\right) +F\left(uf'-2f \right) +f\left(uF'-4F \right) \Biggl. \Biggr]
   +\frac{2 i E \omega  u^4 h'}{F},
\end{autobreak}
\\
\begin{autobreak}
	\tilde{A}=
	\frac{u f h' \theta '}{F}\left( uF'-8F \right),
\end{autobreak}
\\
\begin{autobreak}
	B=
	\frac{u^2 f h'}{F} \left[ u\theta' \left( u f'-6 f \right)+6 \tan \theta \right],
\end{autobreak}
\\
\begin{autobreak}
	\tilde{B}=
	\frac{1}{2 F u f} \Biggl[ \Biggr. u^{2}f^{2}\theta'^{2} \left( uF'-8F \right) +3u^{3}f^{2}h'^{2}\left( uf'-6f\right) +3F\left(uf'-2f\right) -f\left( uF'-4F\right) \Biggl. \Biggr]
	+\frac{2 i E \omega  u^4 h'}{F},
\end{autobreak}
\\
\begin{autobreak}
	C=
	-\frac{k^2}{F f} \Bigl[ u^{4}f^{2}h'^{2}+ F\left(u^{2}f\theta'^{2} +1\right) \Bigr]
	+\frac{3 \sec^{2} \theta}{F u^2 f} \Bigl[ u^{4}f^{2}h'^{2}+ F\left(u^{2}f\theta'^{2} +1\right) \Bigr]
   +\frac{\omega ^2}{F f} \Bigl[ u^{4}fh'^{2} +\left(B^{2}u^{4}+1 \right)\left(u^{2}f\theta'^{2}+1 \right) \Bigr]
   -\frac{i E \omega  u^3 h'}{F f} \Bigl[ uf' +4u^{2}f^{2}\theta'^{2}-2f\left(3u\theta' \tan\theta  +1\right) \Bigr],
\end{autobreak}
\\
\begin{autobreak}
	\tilde{C} =
	-\frac{i E \omega  u \theta '}{F f} \left(4 u^4 f^2 h'^2+uF'-4F \right),
\end{autobreak}
\\
\begin{autobreak}
	D=	
	\frac{i E \omega  u^3}{F^2} \Biggl[ \Biggr. u^{2}f\theta'^{3}\left( uf'-4f \right) +u^{4}f\theta' h'^{2} \left( uf'-6f \right) +6u^{3}fh'^{2}\tan \theta +\theta' \left(uF'-4F \right) \Biggl. \Biggr],
\end{autobreak}
\\
\begin{autobreak}
	\tilde{D} =
	-\frac{k^2}{F f} \Bigl[ u^{4}f^{2}h'^{2}+ F\left(u^{2}f\theta'^{2} +1\right) \Bigr]
	+\frac{3 \sec^{2} \theta}{F u^2 f} \Bigl[ u^{4}f^{2}h'^{2}+ F\left(u^{2}f\theta'^{2} +1\right) \Bigr]
	+\frac{i E \omega  u^3 h'}{F^2} \Bigl[ u^{4}f h'^{2} \left(uf'-6f \right) +\left(u^{2}f \theta'^{2}-1 \right)\left( uF'-4F \right) \Bigr]
  +\frac{\omega ^2}{fF} \Bigl[ u^{4}f h'^{2} +\left( B^{2}u^{4}+1 \right)\left(u^{2}f \theta'^{2}+1 \right) \Bigr].
\end{autobreak}
\end{align}
\end{widetext}

To solve the above equations, we have to impose the appropriate boundary conditions. At the boundary, we impose the condition that the non-normalizable modes of the fluctuations vanish.
At the effective horizon, we impose the ingoing-wave boundary condition.
Here, we write the perturbations near the effective horizon as
\begin{equation}
	\vartheta(u) = (u-u_{*})^{\lambda_{\vartheta}} \vartheta_{\rm reg}(u),
	\:\:\:\:
	a(u) = (u-u_{*})^{\lambda_{a}} a_{\rm reg}(u),\\
\end{equation}
where $\vartheta_{\rm reg}$ and $a_{\rm reg}$ are the regular part of the perturbations at the effective horizon. One finds that the ingoing-wave boundary condition corresponds to $\lambda_{\vartheta}=\lambda_{a}=0$ in our setup. This can be confirmed by introducing the tortoise coordinates for $(t,u)$. The detailed discussion of the ingoing-wave condition in NESS is presented in\,\cite{Mas:2009wf,Ishigaki:2020coe,Ishigaki:2020vtr,Ishigaki:2021vyv}.

To perform the numerical calculations, we discretize the equations of motion with the Chebyshev pseudospectral method. In our study, we used 150 Chebyshev modes along the $u$ direction. 
For the determination of the Chebyshev coefficients, we have employed the Newton-Raphson relaxation scheme within $10^{-7}$ of the root-mean-square error.

\onecolumngrid
\section{Critical exponent \texorpdfstring{$\delta$}{TEXT}}
\begin{figure}[t]
\centering
\includegraphics[width=5.5cm]{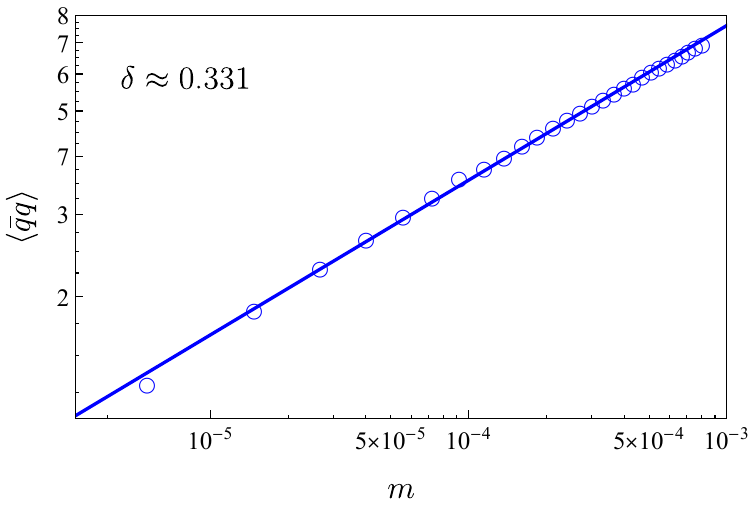}
\includegraphics[width=5.23cm]{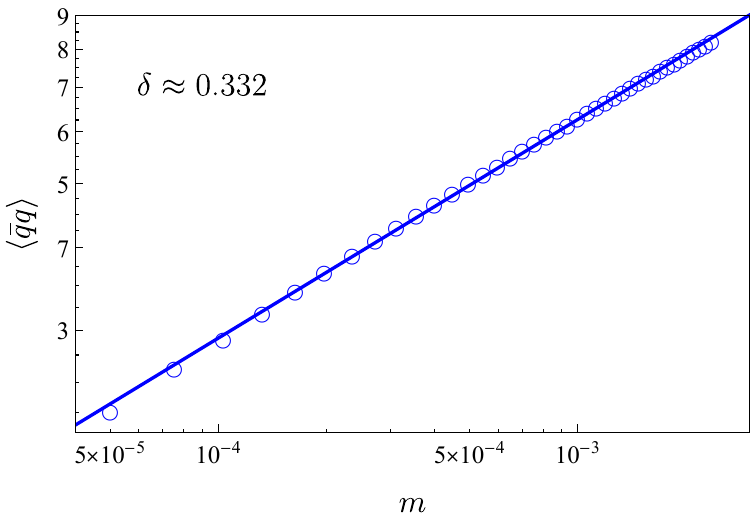}
\includegraphics[width=5.7cm]{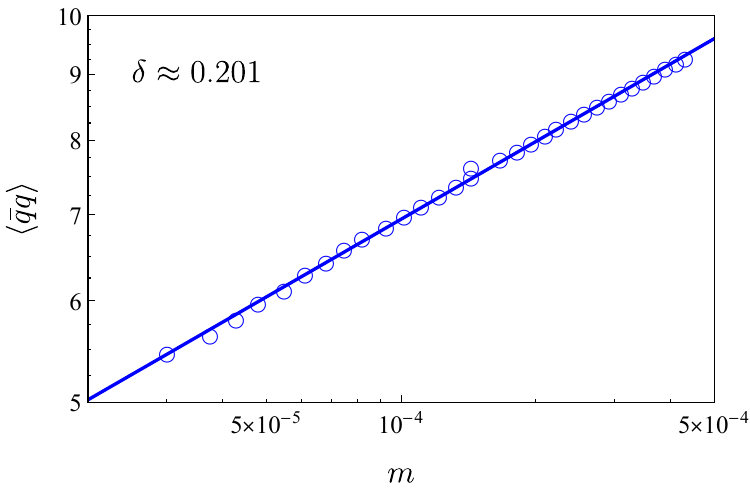}
\caption{The numerical plots of the chiral condensate $\expval{\bar{q}q}$ with respect to $m$ in logarithmic scale. The solid lines denote the results of the linear fitting. The left and middle panels are evaluated at the CL: $T/B^{1/2}\approx 0.096$ and $T/B^{1/2}\approx 0.098$, respectively. The right panel is evaluated at the TCP: $T/B^{1/2}\approx 0.100$. }
\label{fig:delta}
\end{figure}

In the main text, we have studied the critical exponents $(\gamma_{\pm}, \nu_{\pm}, \eta,z)$. In our system, other exponents $(\beta,\delta)$ are defined by
\begin{eqnarray}
    &&\expval{\bar{q}q} \propto \abs{J-J_{c}}^{\beta}, \\
    &&\expval{\bar{q}q} \propto m^{1/\delta}.
\end{eqnarray}
Here, $\delta$ must be evaluated at the critical point. We do not discuss $\alpha$ since the definition of the heat capacity of a NESS is subtle. In the previous paper\,\cite{Imaizumi:2019byu}, we numerically confirmed that $\beta=1/2$ at the CL and $\beta=1/4$ at the TCP. Here, we calculate the value of $\delta$ at the CL and the TCP.  Fig.\,\ref{fig:delta} shows the behavior of the chiral condensate as a function of $m$ at the CL and the TCP. 
The numerical results show that $\delta=3$ at the CL and $\delta=5$ at the TCP. These values agree with those of the Landau theory.

\section{Critical behaviors in other regions}
\begin{figure}[tbp]
\centering
\includegraphics[width=5.5cm]{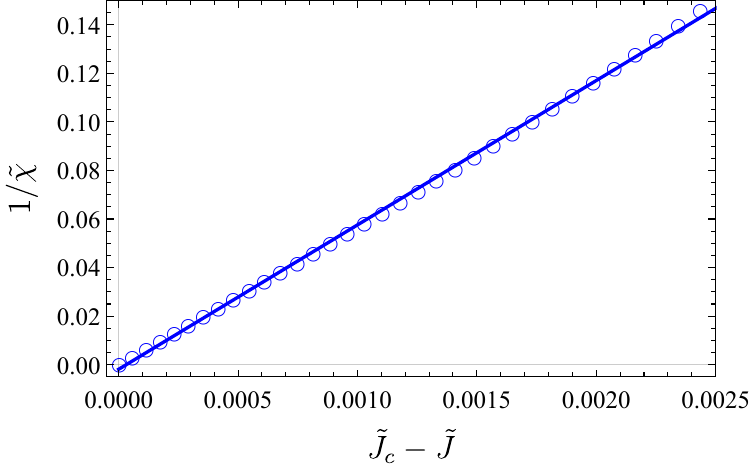}
\includegraphics[width=5.5cm]{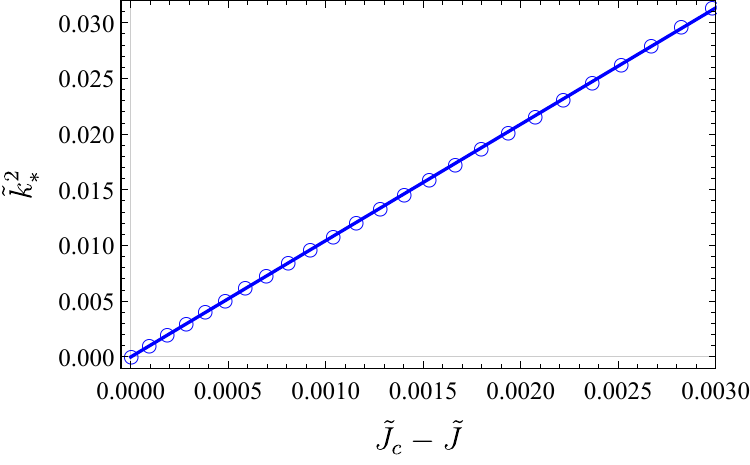}
\includegraphics[width=5.5cm]{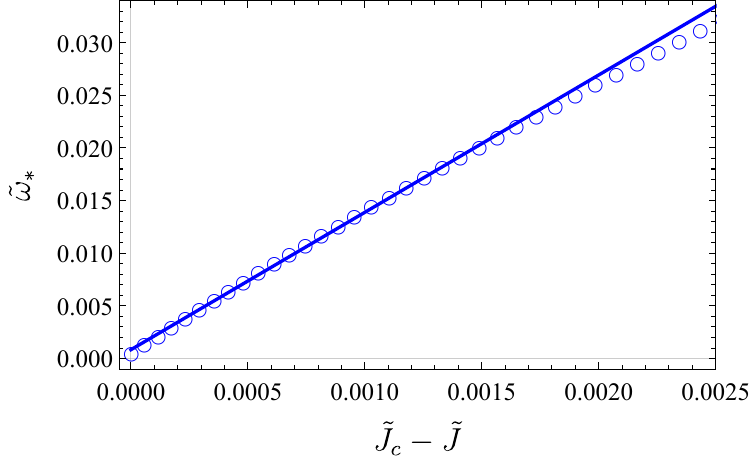}
\caption{The critical behaviors of $1/\tilde{\chi}$
\,(left), $\tilde{k}^{2}_{*}$
\,(middle), and $\tilde{\omega_{*}}$
\,(right) at $T/B^{1/2}\approx 0.096$
in the $\chi {\rm SB}$ phase. Here, the tilde indicates the scaled quantities,
and $\tilde{k}_{*}$, $\tilde{\omega}_{*}$ are the locations of the pole as we have defined in the main text.}
\label{fig:CP2}
\end{figure}

In Fig.\,{\color{blue}2}, we have presented the critical behaviors in the $\chi{\rm SR}$ phase at the CL. The critical behaviors in the $\chi{\rm SB}$ phase at the CL are shown in Fig.\,\ref{fig:CP2}. The critical behaviors in the $\chi{\rm SB}$ phase are qualitatively the same as those in the $\chi{\rm SR}$ phase. Thus, these results give the same values of the critical exponents. 

\begin{figure}[tbp]
\centering
\includegraphics[width=6cm]{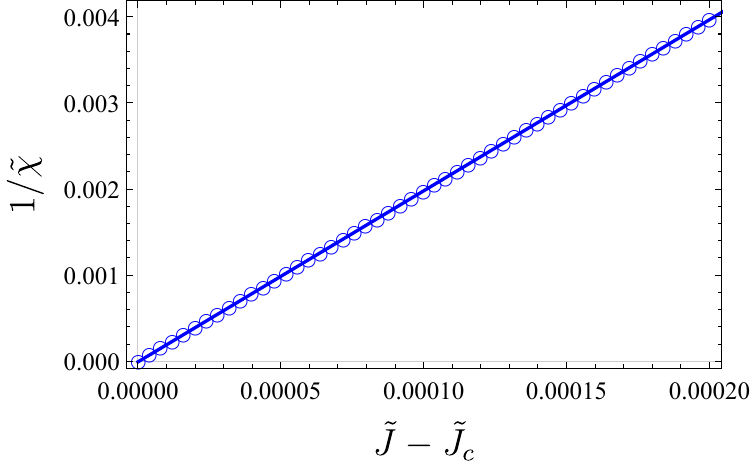}
\includegraphics[width=6cm]{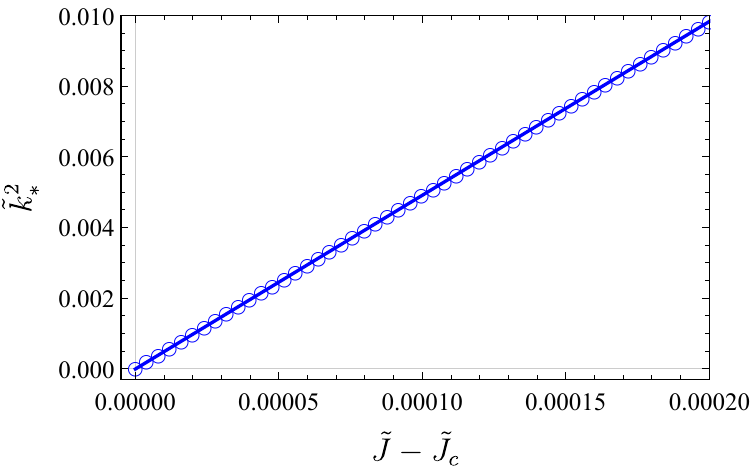}
\includegraphics[width=6cm]{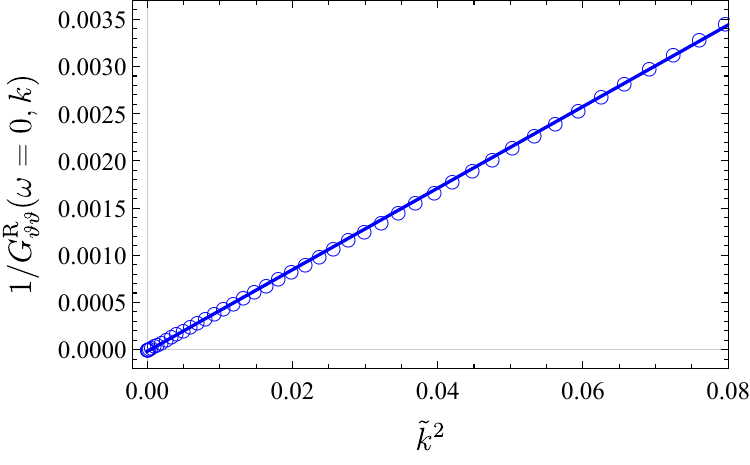}
\includegraphics[width=6cm]{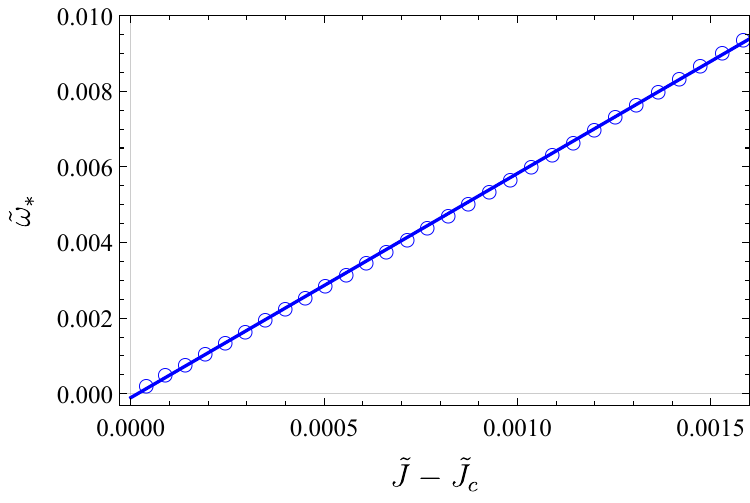}
\caption{The critical behaviors of $1/\tilde{\chi}$
\,(top, left), $\tilde{k}^{2}_{*}$
\,(top, right), 
$G^{\rm R}_{\vartheta \vartheta}(\omega=0,k)$
\,(bottom, left), and $\tilde{\omega_{*}}$ as a function of $\tilde{J}-\tilde{J}_{c}$\,(bottom, right) at $T/B^{1/2}\approx 0.096$
in the $\chi {\rm SB}$ phase. Here, the tilde indicates the scaled quantities
and $\tilde{k}_{*}$, $\tilde{\omega}_{*}$ are the locations of the pole as we have defined in the main text.}
\label{fig:TCP2}
\end{figure}

\begin{figure}[t!bp]
\centering
\includegraphics[width=7cm]{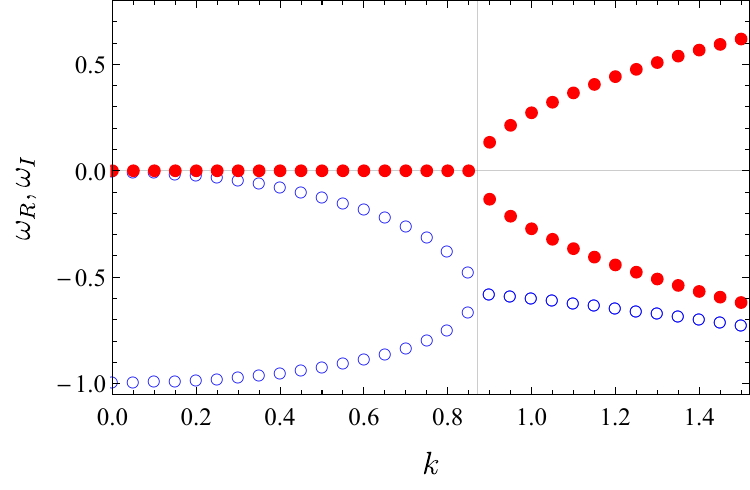}
\caption{The dispersion relation of two different quasi-normal modes at the TCP. The red filled circles and the blue open circles denote numerical plots of the real part and the imaginary part of the quasi-normal frequency, respectively.}
\label{fig:disp}
\end{figure}

Now, we study the critical behaviors in the $\chi{\rm SR}$ phase at the TCP. Fig.\,\ref{fig:TCP2} shows the critical behaviors in the $\chi{\rm SR}$ phase at the TCP. 
We obtain the mean-field values for the critical exponents including $(\gamma_{+},\nu_{+})$.
In the main text, we obtain the value of the dynamic critical exponent $z=2$ from the dynamic scaling analysis. We also confirm this by calculating the dispersion relation at the TCP, which is the so-called critical dispersion relation. At a CL, the dispersion relation is given by $\omega \propto k^{z}$\,\cite{Tuber2014CriticalDA}. In Fig.\,\ref{fig:disp}, we plot the dispersion relation of two different quasi-normal modes. The hydrodynamic mode represents the critical dispersion relation.
As can be seen from Fig.\,\ref{fig:disp}, we find that the critical dispersion relation is that of a diffusive mode:\,$\omega =-i Dk^{2}$, where $D$ is a constant. This indicates $z=2$.
These two modes are pure-imaginary at small momentum. As the momentum increases, however, these pure-imaginary modes get close to each other, and the real parts of them become finite after these modes collide.

In summary, the critical exponents in our phase transition are given in Table\,\ref{table:nonlin}.
Here, $+$ and $-$ represent the values obtained in the $\chi{\rm SR}$ phase and the $\chi{\rm SB}$ phase, respectively. At the CL, all values of the critical exponents agree with those in the mean-field theory. At the TCP, on the other hand, $(\gamma_{\pm}, \nu_{\pm})$ are asymmetric between the $\chi{\rm SR}$ phase and the $\chi{\rm SB}$ phase. Note that these values satisfy the scaling relation for the Green's function:\,$\gamma_{\pm}=\nu_{\pm}(2-\eta)$.
\begin{table}[tbp]
\caption{Critical exponents at the CL and the TCP} 
\centering 
\begin{tabular}{c c c} 
\hline  
 \hspace{3em} & \hspace{2em}CL\hspace{2em} & \hspace{2em}TCP\hspace{2em} \\ [0.5ex] 
\hline 
$\beta$ & 0.5 & 0.25  \\ 
$\delta$ & 3 & 5  \\
$\gamma_{\pm}$ & 1\,($\pm$) & 0.58\,($-$), 1\,($+$) \\
$\nu_{\pm}$ & 0.5\,($\pm$) & 0.29\,($-$), 0.5\,($+$) \\
$\eta$ & 0 & 0  \\
$z$ & 2 & 2\\[0.5ex] 
\hline 
\end{tabular}
\label{table:nonlin} 
\end{table}

\bibliography{main}

\end{document}